\documentclass[a4paper,11pt]{article}

\usepackage{latexsym}
\usepackage{amsmath}
\usepackage[dvips]{graphicx}

\title{Unification, Multiplets and Proton Decay}

\author{
        Tuhin Roy\, \thanks{\tt tuhin@physics.bu.edu}\ \
        \\ 
        \small \sl \ Physics Department, Boston University, 
                Boston, MA  02215\\ \\ 
       }

\newcommand{\beq}{\begin{eqnarray}}
\newcommand{\eeq}{\end{eqnarray}}

\newcommand{\al}{\alpha}
\newcommand{\De}{\Delta}
\newcommand{\de}{\delta}
\newcommand{\be}{\beta}

\newcommand{\drawsquare}[2]{\hbox{%
\rule{#2pt}{#1pt}\hskip-#2pt
\rule{#1pt}{#2pt}\hskip-#1pt
\rule[#1pt]{#1pt}{#2pt}}\rule[#1pt]{#2pt}{#2pt}\hskip-#2pt
\rule{#2pt}{#1pt}}

\newcommand{\Yfund}{\raisebox{-.5pt}{\drawsquare{6.5}{0.4}}}
\newcommand{\Ysymm}{\raisebox{-.5pt}{\drawsquare{6.5}{0.4}}\hskip-0.4pt%
        \raisebox{-.5pt}{\drawsquare{6.5}{0.4}}}
\newcommand{\Ythrees}{\raisebox{-.5pt}{\drawsquare{6.5}{0.4}}\hskip-0.4pt%
          \raisebox{-.5pt}{\drawsquare{6.5}{0.4}}\hskip-0.4pt%
          \raisebox{-.5pt}{\drawsquare{6.5}{0.4}}}

\newcommand{\Yadjoint}{\raisebox{-3.5pt}{\drawsquare{6.5}{0.4}}\hskip-6.9pt%
        \raisebox{3pt}{\drawsquare{6.5}{0.4}}\hskip-0.4pt
        \raisebox{3pt}{\drawsquare{6.5}{0.4}}}
\begin{document}
\baselineskip=17pt 
\pagestyle{plain}

\begin{titlepage} 
\vskip-.4in 
\maketitle 
\begin{picture}(0,0)(0,0) 
\put(360,200){BUHEP-04-14}
\end{picture}

\begin{abstract} 
\leftskip-.6in 
\rightskip-.6in 
\vskip.4in 

We make a detailed analysis of  gauge coupling unification in
supersymmetry. When the Standard Model gauge group is embedded in a 
Grand Unified Theory, new particles often appear below the GUT scale in 
order to predict the right
phenomenology at low energy. While these new particles are beyond the reach of
accelerator experiments, they change the prediction of $\alpha_s$. Here we
classify all the $(\mathrm{SU(3)_C,SU(2)_L)_{U(1)_Y}}$ 
representations which improve or worsen the
prediction. Running experimentally determined values of the coupling
constants at two loops we calculate the allowed range of masses of fields in
these 
representations. We explore the implication of these results in SU(5) and
$\mathrm{SU}(3)^3$ (trinification) models. We discover that minimal
trinification predicts light triplet Higgs particles which lead to proton
decay with a lifetime in the vicinity of the current experimental bound.

\end{abstract} 
\thispagestyle{empty} 
\setcounter{page}{0} 
\end{titlepage} 
\section{Introduction}

The Standard Model of particle physics provides
a good description of almost all non-gravitational phenomena of known
particles and their strong and electroweak interactions. 
Yet, its failure to address many theoretical issues has motivated
a search for supersymmetry. Low energy supersymmetry is  the most popular 
solution to the hierarchy problem.  Experimentally,
the best indication  of supersymmetry is the prediction of gauge coupling
unification in the simplest supersymmetric extension of the Standard Model
({\it viz.} MSSM).  Quantitatively, the coupling constants unify in the MSSM
at 99.73\% confidence level, corresponding to 
$3\sigma$ for one degree of freedom.

From a top-down point of view, specific GUT models generically predict new
particles beyond the MSSM. They change the scale dependence of the coupling
constants. If all  new particles are at  the GUT scale or if they form
complete multiplets of a unifying group, then the
gauge coupling unification is as automatic as in the MSSM.  However, 
to predict low-energy physics correctly few
$(\mathrm{SU(3)_C,SU(2)_L)_{U(1)_Y}}$  multiplets usually become light. To be
more  explicit, let us point out a few reasons
that may result in light degrees of freedom in model building
\begin{enumerate}
\item New particles may get masses from higher dimensional operators. In this
  case their mass is suppressed by  $\Lambda$, the cut-off scale of the
  theory. 
\item Often these new particles appear in the  representation that contains
  quarks and leptons. In that case their mass may become proportional to the
  Yukawa  couplings in the MSSM. Thus, the tiny masses of quarks and
  leptons in the first two generations may make these new particles lighter
  than the GUT scale
\item Light particles may arise as pseudo Goldstone bosons of spontaneously
  broken approximate global symmetries of
  the theory. Also, some new particles may become light in the absence of
  specific superpotential couplings. SUSY ensures that these couplings will
  not be generated by loops.  
\end{enumerate}

The object of this paper is to present a systematic way to talk  about 
so called ``GUT-scale threshold effects'' due to these new light particles
in a model-independent way. Note that there is another kind of the threshold
corrections at the GUT scale because of higher dimensional operators
suppressed by the Planck mass. These operators can produce significant
corrections to the gauge kinetic function. For a recent discussion on
effects of these  operators on unification and references
see~\cite{Tobe:2003yj}. 
In this paper we ignore these contributions of non-renormalizable
operators. We 
demand that gauge coupling constants unify within 99.73\% confidence
level as in the MSSM even with new particles at intermediate mass
scales. This allows us 
to isolate all  $(\mathrm{SU(3)_C,SU(2)_L)_{U(1)_Y}}$ 
multiplets which, when light, push the
coupling constants away from each other and worsen the prediction of
$\alpha_s$. For example, electroweak doublets are in the category of 
{\bf bad} particles which worsen  unification when they become light.   
We also take the experimentally determined values of the coupling
constants as input and use the constraint of unification to
predict the range of masses of different representations.   
Constraints from unification have previously been used to estimate the mass of
the colored Higgs in the SU(5) model 
~\cite{Hisano:1992mh}\cite{Hisano:1994hb}\cite{Hisano:1997nu}\cite{Murayama:2001ur}.    
Here we present the results for all low dimensional 
 $(\mathrm{SU(3)_C,SU(2)_L)_{U(1)_Y}}$
multiplets, irrespective of any underlying gauge symmetry in a renormalizable
theory.
We apply the formalism developed here to SU(5) and $\mathrm{SU}(3)^3$
Grand Unified Theory. In particular, we estimate the size of 
doublet-triplet splitting in the {\bf 5} of SU(5).

In Section 3 we perform a detail study of the $\mathrm{SU}(3)^3$ 
trinification scheme~\cite{Glashow:1984gc}\cite{Lazarides:1994uw}
\cite{Dvali:1994wj}\cite{Dvali:1994vj}\cite{Dvali:1996fc}\cite{Willenbrock:2003ca}
\cite{Kim:2004pe}.    
We construct a phenomenologically viable minimal trinification model that
also preserves unification. We find that minimal trinification is not
absolutely safe from proton decay constraints. In trinification specific
models of Yukawa matrices predict proton decay, mediated by the scalar part of
the colored Higgs, with a lifetime significantly shorter than that of SU(5)
GUTs. In minimal trinification the constraint of unification lowers the
mass of colored Higgs to $10^{14}\mathrm{GeV}$. We show here that 
in these specific models, proton decay become interesting
because of smaller colored Higgs mass. We also propose here a simple
extension of the {\it minimal} model, without enlarging the number of
$\mathrm{SU}(3)^3$ multiplets, that can avoid the difficulties of  
{\it  minimal} trinification.

\section{Unification}

We start by showing that gauge coupling constants unify in the MSSM with present 
day data and uncertainties. Next we introduce
different multiplets close to the GUT scale and  investigate how the constraint
of exact unification can be implemented to yield new insights 
into the picture of unified theories. 

\subsection{Unification in the MSSM}
We have carried out numerical calculations for the two loop RGEs of the
gauge and Yukawa coupling constants from the SUSY scale $M_S$ to the GUT
scale $M_G$. RGEs for both the gauge couplings
and the Yukawa couplings are well documented ~\cite{Barger:1993ac}. 
We take $M_S$ as 1 TeV and include only the Yukawa coupling of the top quark.

The prediction of unification is extremely sensitive to the used
value of $\al_s(m_Z)$~\cite{Dorsner:2003yg}. Note that there are large
discrepancies in the values of the strong coupling constant determined from
high energy experiments and low energy experiments (especially $\Upsilon$
decay)\footnote{see QCD section of ~\cite{Eidelman:2004wy} for discussion}.
PDG quotes the average value as $\al_s(m_z) = 0.1187 \pm
0.002$~\cite{Eidelman:2004wy}. 
However, the  global fit to precision electroweak analysis\footnote{strictly,
  one should use MSSM fit to the electroweak data. However, $\al_s$ changes
  only by fractions of $\sigma$ when fitted with MSSM parameters compared to 
  the results of SM fit (see \cite{deBoer:2003xm}  for details).} generates a higher
value ($\al_s(m_z) = 0.1213 \pm 0.0018$)~\cite{Eidelman-Langacker:2004wy}.
There are also SUSY threshold contributions from the unknown sparticle spectrum. 
In order to determine the sensitivity of our results to the values of 
 $\al_s$ as well as the gaugino/squark/slepton/Higgs and Higgsino masses we
 performed our analysis for a series of different input parameters.

We begin with the assumption that all the sparticles other than the gauginos,
appear at the scale $M_S$. We take winos at 200GeV and the gluinos at
700GeV. We use the following precision measurements as 
inputs~\cite{Eidelman:2004wy}
\beq
\al_{s_{\overline{MS}}}(m_z) = 0.1187 \pm 0.002 ,\nonumber\\
\mathrm{sin}^2\theta_{w_{\overline{MS}}}(m_z) = 0.23120 \pm 0.00015 ,\nonumber\\
\al_{{em}_{\overline{MS}}}^{-1}(m_z) = 127.906 \pm 0.019  .
\label{input}
\eeq    
These quantities are given in the $\overline{MS}$
scheme. We use  reference in~\cite{Martin:1993yx} to convert them 
to the $\overline{DR}$ scheme, as only in the $\overline{DR}$ scheme that can 
we approximate the RGEs as  step functions  at the particle threshold 
~\cite{Antoniadis:1982vr}.

\begin{figure}[t]
\begin{center}
\includegraphics[width=0.7\textwidth]{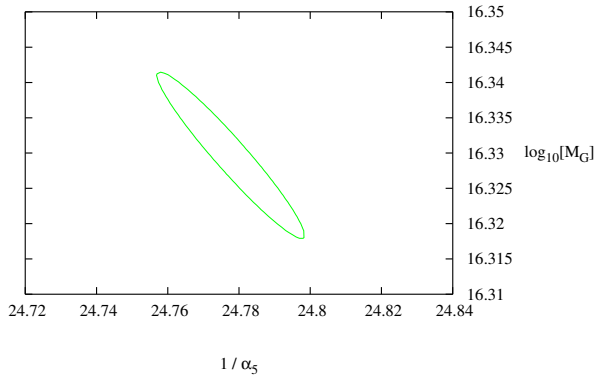}
\end{center}
\caption{Plot showing 99.73\% contour for two loop SUSY running with only
  the MSSM particle content. $\al_5$ is the unified coupling constant at
  the GUT scale.} 
\label{mssmchi}
\end{figure}

Operationally we use  given values of the unified coupling constant $\al_5$
and $M_G$ to predict the data of Eq.(\ref{input}). We find that 
gauge coupling constants unify with present data (using the $\chi^2$ fit
for two degrees of freedom) at the 99.73\% confidence level for all values of
$\al_5$ and $M_G$ which lie in the ellipse shown in Fig.(\ref {mssmchi}). 

However, a slightly higher value of $\al_s$ improves unification
significantly. If we use $\al_s(m_z) = 0.1213 \pm 0.0018$ as input parameter 
instead, coupling constants unify at 95\% confidence level.

Whether gauge couplings unify within error bars or not   thus
depends on the  value of $\al_s$ one decides to use.  For example,
if one uses the PDG average, coupling constants do
not unify within  95\% confidence level. In minimal supersymmetric 
SU(5) model, as a result, a  GUT threshold effect from the light colored
Higgs is needed  in order to
predict  $\al_s$ correctly~\cite{Murayama:2001ur}. The light colored Higgs,
in turn, accelerates proton decay. Detailed discussions of proton decay in
minimal SU(5) model can be found in  
~\cite{Hisano:1992mh}\cite{Hisano:1994hb}\cite{Hisano:1997nu}\cite{Murayama:2001ur} 
\cite{Bajc:2002bv}\cite{Bajc:2002pg}.
On the other hand, if we choose the result of the global electroweak fit
instead, we find that  the coupling constants unify at 95\% confidence level
and no GUT threshold effect is needed.

To check the effects of sparticles spectrum we also perform the same
calculation after introducing   splittings between sleptons and
squarks. Note that the overall scale for quarks and
sleptons does not have significant effects in the running, because: (a) they
come in complete multiplets of SU(5) and do not affect unification of
couplings at 1-loop. Only squark-slepton mass splittings produce changes. 
(b) Squarks and sleptons have smaller contributions to the $\be$ functions, 
being scalar
parts of super-multiplets. We find that the contributions of a large splitting 
({\it viz.} $m_{\tilde{sq}}/m_{\tilde{sl}}=(M_3/M_2)^2$) are equivalent to
shifting the input value of $\al_s$ by $1\sigma$. We also assume that the
heavy Higgs scalar mass ($m_H$) and the Higgsino mass ($\mu$) are at
$1\,$TeV. The 
combined threshold corrections of heavy Higgs and Higgsinos are maximum if
both $m_H$ and $\mu$ are lighter (or heavier) than $1\,$TeV. If both 
$m_H$ and $\mu$ are assumed to be 500 GeV, we again find effects which are
slightly less than the contributions of shifting the input value of $\al_s$
by $1\sigma$. To find the effect of different gaugino masses, we  have also
varied $M_2$ from 100 GeV to 600 GeV keeping $M_3/M_2$ fixed. 
This shift of gaugino masses also changes the prediction of $\al_s$
by $1\sigma$.  

Currently,  $\Upsilon$ decay measures a  low value of $\al_s$ (more
than $2\sigma$ smaller than the PDG average). Whereas, the global electroweak
fit results in a high $\al_s$ (more than $1\sigma$ bigger than the 
PDG average). We also find that the fluctuation of sparticle spectrum can
produce an effect,  equivalent to shifting the central value of 
$\al_s$ by  $1\sigma$. Given the sensitivity of gauge coupling unification on
the input value of  $\al_s$, we take a conservative and simplified
approach. We choose the PDG average along with the quoted error
(Eq.(\ref{input})) as input. However we present all our results at 99.73\% 
confidence level, {\it i.e.} we basically allow $3\sigma$ variation of  
$\al_s$ in our calculation. Gauge coupling
constants unify within 99.73\% confidence level and therefore we demand that
unification is maintained at this level of accuracy when contributions of new
physics are added to the RGEs. 
We use the sparticle spectrum as was described in the beginning of
this section ({\it i.e.} $M_2=200\,$GeV and all extra particles in the MSSM other
than the gauginos, are at $1\,$TeV). Also, allowing  $3\sigma$ error in the
predicted value of $\al_s$ reduces the sensitivity to SUSY threshold effects.

\subsection{Unification with an extra multiplet}

New particles are expected to appear in various GUT models near the
unification scale. If they become lighter than the
unification scale, the scale dependence of the coupling constants gets
modified. In this section, we discuss the effects of new multiplets in a
model-independent way. However,  as we already have pointed out in the
introduction, we neglect contributions from higher dimensional operators.

We insert a general multiplet $(a,b)_Y$ along with its adjoint and give it a
Dirac mass $M_X$.
Here $a$ and $b$ are the dimensions of the representations of the multiplet
under $\mathrm{SU(3)_C}$ and $\mathrm{SU(2)_L}$, and $Y$ is the hypercharge.
This insertion is either going to push the coupling constants away from each
other, making the unification worse compared to the case of only MSSM particle
content, or it works toward having a better unification. We call all these
multiplets whose insertion at a scale lower than the unification scale worsen
unification prediction, ``{\bf bad}'' multiplets, while 
other multiplets will be called ``{\bf good}'' multiplets. 

Now, what makes $(a,b)_Y$  a  {\it  bad} multiplet? 
By definition a {\it bad} multiplet worsens unification.  Hence it should
be clear that its insertion can never allow the central values of all the
three coupling constants to coincide at a point, as they don't even do so
in the MSSM. We define the  {\it bad} multiplets as the ones for which
there exists no scale of insertion  $M_X$ that produces exact unification.
In order to turn this definition into a constraint on the
representation $(a,b)_Y$ we
impose the condition that the central values of the three coupling constants 
coincide at the unification scale. 
For {\it bad} multiplets this condition can never 
be satisfied in the acceptable region of $M_X < M_G$.       

To begin, we introduce new variables for the differences of coupling
constants. They are more suitable when we talk about unification
\begin{eqnarray}
\De_{ij}(M)\equiv \frac{1}{\al_i}(M) - \frac{1}{\al_j}(M) .
\end{eqnarray}
The virtue of using this alternative language may be demonstrated by
inserting a {\bf 5} of SU(5). Although the coupling constants themselves
change because of this insertion, $\De_{ij}$ remain  unaltered,
reflecting the fact that a  {\bf 5}  of SU(5) does not change unification at
one loop. 

To distinguish the situation before and after the insertion of the extra
multiplet, the coupling constants and the difference variables found in
the MSSM  are denoted as $\al_i^0$ and $\De_{ij}^0$.   

The introduction of a multiplet at scale $M_X$ changes the running  
\beq
\frac{1}{\al_i}(M)=\frac{1}{\al_i^0}(M)-\frac{\de  \be_i}{2\pi}
                          \ln (M/M_X) \label{eq:mx} .
\eeq
where $\al_i^0(M)$ is the value as determined in the MSSM at two loops and
$\de\be_i$ is the contribution of the new multiplet to the corresponding
$\be$ function. Consequently, the values of $\De_{ij}$ are also shifted 
\beq
\De_{ij}(M)= \De_{ij}^0(M)-
     \frac{1}{2\pi}(\de \be_i-\de \be_j)\ln(M/M_X) \label{eq:Dmx} .  
\eeq
We impose the condition
\beq
 \al_5 =\al_1(M_G)=\al_2(M_G)=\al_3(M_G)  \label{constraint1} .
\eeq
Which implies
\beq
 \De_{13}(M_G)=\De_{12}(M_G)=0 \label{constraint2} .
\eeq 
Eqs.~(\ref{constraint1}) and (\ref{constraint2}) determine $M_G$.
After using (\ref{eq:Dmx}) we find that at the GUT scale
\beq
\De_{12}^0(M_{G})&=&
     \frac{1}{2\pi}(\de \be_1-\de \be_2)\ln(M_G/M_X)  ,\nonumber\\
\De_{13}^0(M_{G})&=&
     \frac{1}{2\pi}(\de \be_1-\de \be_3)\ln(M_G/M_X)  .
\label{eq:condn} 
\eeq
These equations can be solved to determine $M_X$ and
$M_G$. However, we are only interested in solutions for which $M_X < M_G$. 
This condition along with Eq.(\ref{eq:condn})
constrains the size of $\de\beta_i$. 

Using numerical values of $\De_{12}^0$ and $\De_{13}^0$ we find that 
all multiplets satisfying $\de\be_2 \geq\mathrm{Max}(\de\be_3,\de\be_1)$ never
satisfy Eq.(\ref{eq:condn}). In the numerical calculation we also use an
additional constraint that $10^{14}\mathrm{GeV}\leq M_G \leq M_{Pl}$.  
A detailed study reveals that all the multiplets for which the solutions to 
Eq.(\ref{eq:condn}) satisfy our constraints can be classified as
\begin{itemize}
 \item {\bf Type 1:}  $\de \be_2 \geq \de \be_3$ and 
   $\de\be_1\geq\de\be_2+\mathrm{Max}(0.2,\frac{3}{2}(\de\be_2-\de\be_3))$.
 \item {\bf Type 2:}  $\de\be_2 <  \de\be_3$ and 
   $\de\be_1\geq\mathrm{Max}(0, \de\be_2-\frac{14}{11}(\de\be_3-\de\be_2))$.    
\end{itemize}
By definition, both Type 1 and Type 2 are  {\it good} multiplets. The factors 
$\frac{3}{2}(\de\be_2-\de\be_3)$ in Type 1 and
$\frac{14}{11}(\de\be_3-\de\be_2)$ in Type 2 are included to ensure that $M_G
> 10^{14}\mathrm{GeV}$ and $M_G \leq 10^{19}\mathrm{GeV}$ respectively. In
case of Type 1 multiplets, if  $\de\be_2 = \de\be_1$, we need $\de\be_1 \geq
\de\be_2+0.2$ so that $M_X$ is greater than 1TeV.   
For any other multiplet, there exists no acceptable set of $M_X$ and $M_G$
for which the central values of the coupling constants coincide.  These are
the  {\it bad} multiplets.

Before proceeding let us discuss the meaning of these classifications. 
\begin{enumerate}
\item  For $\de \be_2 \geq \de \be_3$, there exists a critical value of $Y$
  (say $Y_{min}$). A multiplet  $(a,b)_Y$ with $Y^2\geq Y_{min}^2$ is Type 1 
  and for $Y^2 < Y_{min}^2$ it is a {\it  bad} multiplet.
  $Y_{min}$ is determined by the size of $a$ and $b$. For example, in the case
  of the representation  $(1,2)_{Y}$, we find $Y_{min}^2$ to be $25/24$. Thus,  
  the multiplet $(1,2)_{1/2}$, which frequently appears in various models,  
  is a {\it bad} multiplet. 
\item Similarly, for Type 2 multiplets there are lower bounds on
  $\de\be_1$ implying   constraints on the sizes of their hypercharge.
\item If there is a Type 1 multiplet in the theory, it lowers the GUT scale
  from the MSSM $M_G$. On the other hand, if it is a Type 2 multiplet with 
  $\de\be_2 > \de\be_1$, $M_G$ goes up. 
\item We used the  central values of the coupling constants to determine
  various  numerical factors in the classification discussed. We use this
  result  to identify different multiplets  
  (which actually appear in GUT models) whether they improve or worsen
  unification.  However dimensions of SU(3) and SU(2) representations can
  only have 
  integer values and consequently the $\beta$ functions are discrete,
  differing by order 1 numbers. Therefore the classification is insensitive
  to small shifts of input values. For the same reason,
  constraints from the condition  $M_X \geq 1\,$TeV are also trivially
  satisfied. 
\end{enumerate}
Table~\ref{Ymax} lists all the low dimensional {\it good} multiplets with
classifications discussed above.
\begin{table}[t]
\begin{tabular}{|c c|c c c||c c|c c c|}
\hline
\multicolumn{5}{|c||}{$\de \be_2 \geq \de \be_3$} 
             & \multicolumn{5}{|c|}{$\de \be_2 < \de \be_3$}\\
\multicolumn{5}{|c||}{Type 1: $Y^2 \geq Y_{min}^2$} 
             & \multicolumn{5}{|c|}{Type 2:  $Y^2 \geq Y_{min}^2$}\\ 
\hline
SU(3) & SU(2)  & $\delta \be_3$ & $\delta \be_2$ & $Y_{min}^2$ &
    SU(3) & SU(2)  & $\delta \be_3$ & $\delta \be_2$ &$Y_{min}^2$\\
\hline
\hline
    1     &    1   & 0  & 0  &    1/6 &   \Yfund &    1  & 1  & 0 &      0\\
    1     &\Yfund  & 0  & 1  &  25/24 & \Yadjoint&    1  & 3  & 0 &      0\\ 
 \Yfund   &\Yfund  & 2  & 3  &    5/8 &  \Ysymm  &    1  & 5  & 0 &      0\\
    1     &\Ysymm  & 0  & 4  &   25/9 & \Ythrees &    1  & 15 & 0 &      0\\
 \Yfund   &\Ysymm  & 3  & 12 &255/108 &  \Ysymm  & \Yfund& 10 & 6 & 25/396\\
 \Ysymm   &\Ysymm  & 15 & 24 & 125/72 & \Yadjoint& \Yfund& 12 & 8 &   5/33\\
 \Yadjoint&\Ysymm  & 18 & 32 &265/144 & \Ythrees & \Yfund& 30 & 10&      0\\
    1     &\Ythrees& 0  & 10 & 125/24 & \Ythrees & \Ysymm& 45 & 40&185/198\\
 \Yfund   &\Ythrees& 4  & 30 & 115/24 &          &       &  &   &       \\
 \Ysymm   &\Ythrees& 20 & 60 &   25/6 &          &       &  &   &       \\
 \Yadjoint&\Ythrees& 24 & 80 & 205/48 &          &       &  &   &   \\
\hline 
\end{tabular}
\caption{Classification of all the low dimensional representations.  
  Multiplet $(a,b)_Y$ with $\de \be_2 \geq \de \be_3$ is a 
  Type 1  {\it good} multiplet if $Y^2\geq Y_{min}^2$ and  {\it bad}
  otherwise.  Similarly, a multiplet with  $\de \be_2 < \de \be_3$ is a 
  Type 2 {\it good} multiplet if its hypercharge is more than corresponding
  $Y_{min}$ as listed here.}
\label{Ymax}
\end{table}
 
At the beginning of this section, we called all multiplets, which improve
unification compared to the MSSM, {\it good} multiplets. To identify them we
put a stronger constraint and redefined the  {\it good} multiplets as the
ones for which the central values of the coupling constants coincide, with
physical conditions $M_S \leq M_X < M_G$ and $10^{14}\mathrm{GeV}\leq M_G 
\leq M_{Pl}$.  
All  multiplets listed in Table~\ref{Ymax} are {\it good} multiplets based
on this redefinition.

\subsection{Various models}

In the last section, we have identified all the small
multiplets, which if appearing in the desert, worsen the degree of
unification  present in the MSSM.  
The next job is to check whether these multiplets  actually appear
in different GUT schemes or not. If they do, we need to make sure that
there are enough light {\it good} multiplets to control the damage. To do
so, we need to pick different types of theories
and look at various multiplets that originate when the bigger group is
decomposed to the MSSM gauge group.    

We now determine which multiplets actually appear in various GUTs by
decomposing the GUT multiplets into  $\mathrm{SU}(3)_\mathrm{C}\times
\mathrm{SU}(2)_\mathrm{L}\times\mathrm{U}(1)_Y$ representations.     
Most common GUTs have gauge groups 
SU(5), SO(10), $ \mathrm{SU(3)}^3$ and $ \mathrm{E}_6$. The MSSM is contained 
in SO(10) through an SU(5)
and in $ \mathrm{E}_6$ through an $ \mathrm{SU(3)}^3$ or  SO(10). 
Therefore we limit our attention to branching rules of SU(5) and $
\mathrm{SU(3)}^3$. 
Eqs.~(\ref{su5sm}) and (\ref{trnsm}) show the branching rules for 
the low dimensional representations of SU(5) and $ \mathrm{SU(3)}^3$
respectively as the group is broken to $\mathrm{SU}(3)_\mathrm{C}\times
\mathrm{SU}(2)_\mathrm{L}\times\mathrm{U}(1)_Y$~\cite{Slansky:1981yr}. 
The {\it bad} multiplets are underlined in both cases.

{\bf Branching Rules for $\mathrm{SU(5)} \supset \mathrm{SU}(3)_ \mathrm{C}
\times \mathrm{SU}(2)_ \mathrm{L} \times\mathrm{U}(1)_Y$. }
\beq
5 &=& \underline{(1,2)_{1/2}}+(3,1)_{-1/3} \nonumber \\
10&=& (1,1)_1+(\bar 3,1)_{-2/3}+\underline{(3,2)_{1/6}} \nonumber\\
15&=& \underline{(1,3)_1}+\underline{(3,2)_{1/6}}+
     (6,1)_{-2/3}\nonumber\\
24&=& (1,1)_0+\underline{(1,3)_0}+(3,2)_{-5/6}+(3,2)_{5/6}
                               +(8,1)_0\nonumber\\
35&=& \underline{(1,4)_{-3/2}}+\underline{(\bar 3,3)_{-2/3}}+
              \underline{(\bar 6,2)_{1/6}}
                    +(\overline{10},1)_{1}\nonumber\\
40&=& (1,2)_{-3/2}+\underline{(3,2)_{1/6}}+(\bar 3,1)_{-2/3}+
      \underline{(\bar 3,3)_{-2/3}}+ (8,1)_1
          +\underline{(\bar 6,2)_{1/6}}\nonumber\\
45&=& \underline{(1,2)_{1/2}}+(3,1)_{-1/3}+\underline{(3,3)_{-1/3}}+
             (\bar 3,1)_{4/3}+(\bar 3,2)_{-7/6}+
           (\bar 6,1)_{-1/3}+(8,2)_{1/2}\nonumber\\
50&=& (1,1)_{-2}+(3,1)_{-1/3}+(\bar 3,2)_{-7/6}+\underline{(\bar 6,3)_{-1/3}}
          +(6,1)_{4/3}+(8,2)_{1/2}
\label{su5sm}
\eeq

{\bf Branching Rules for $\mathrm{SU}(3)_{\mathrm{C}} \times  \mathrm{SU}(3)_
{\mathrm{L}} \times  \mathrm{SU}(3)_{\mathrm{R}} \supset \mathrm{SU}(3)_
{\mathrm{C}}  \times \mathrm{SU}(2)_{\mathrm{L}}\times\mathrm{U}(1)_Y$}
\beq
(3,\bar 3,1)&=& \underline{(3,2)_{1/6}} + (3,1)_{-1/3} \nonumber\\
(\bar 3,1,3)&=& (\bar 3,1)_{1/3}+(\bar 3,1)_{1/3}+
                              (\bar 3,1)_{-2/3} \nonumber\\
(1,3,\bar 3)&=& \underline{(1,2)_{1/2}}+\underline{(1,2)_{-1/2}}+
                \underline{(1,2)_{-1/2}}+(1,1)_1+(1,1)_0 +(1,1)_0 \nonumber\\
(1,1,8)&=& (1,1)_1+(1,1)_0+(1,1)_0+(1,1)_{-1}+(1,1)_1+(1,1)_0+(1,1)_{-1}+
                (1,1)_0 \nonumber\\
(1,8,1)&=& \underline{(1,2)_{1/2}}+\underline{(1,2)_{-1/2}}+
             \underline{(1,3)_0}+\underline{(1,1)_0} \nonumber \\
(8,1,1)&=&(8,1)_0
\label{trnsm}
\eeq

Eqs.~(\ref{su5sm}) and (\ref{trnsm}) also find {\it good} multiplets, which
appear in different GUT schemes. If these multiplets 
become lighter than the GUT scale they push the coupling constants
toward each other and for some value of $M_X$ we predict the central 
values to meet exactly at a point. But a {\it good} multiplet can also worsen
unification if it becomes too light. Thus we have the lower
bound on the mass  $(M_X^{min})$ above which the coupling constants unify
within 99.73\% confidence level. We  
have carried out our investigation for all low dimensional {\it good}
multiplets that appear in Eqs.~(\ref{su5sm}) and (\ref{trnsm}) and found the 
values of $(M_X^{min})$ from the current data. These are listed in the
Table~\ref{mx}.

\begin{table}[t]
\label{Mgut}
\begin{tabular}{|l|c|c|}
\hline
multiplets & $M_{GUT}$ & $M_X^{min}$  \\
\hline
\hline
$(1,1)_{1}$   &$5.1\times 10^ {15}-2.1\times 10^ {16}$&$6.2\times 10^ {12}$\\
$(1,1)_{-2}$  &$4.8\times 10^ {15}-2.1\times 10^ {16}$&$9.6\times 10^ {14}$\\
$(1,2)_{3/2}$ &$2.7\times 10^ {15}-2.0\times 10^ {16}$&$1.9\times 10^ {14}$\\
$(3,1)_{1/3}$ &$1.5\times 10^ {16}-2.2\times 10^ {16}$&$1.6\times 10^ {14}$\\
$(3,1)_{2/3}$ &$1.0\times 10^ {16}-2.1\times 10^ {16}$&$6.9\times 10^ {14}$\\
$(3,1)_{4/3}$ &$6.6\times 10^ {15}-2.1\times 10^ {16}$&$2.3\times 10^ {15}$\\
$(3,2)_{5/6}$ &$2.2\times 10^ {14}-1.7\times 10^ {16}$&$5.6\times 10^ {8}$\\
$(3,2)_{7/6}$ &$3.7\times 10^ {15}-2.0\times 10^ {16}$&$8.9\times 10^ {14}$\\
$(6,1)_{1/3}$ &$1.8\times 10^ {16}-2.3\times 10^ {16}$&$6.5\times 10^ {15}$\\
$(6,1)_{2/3}$ &$1.4\times 10^ {16}-2.2\times 10^ {16}$&$6.2\times 10^ {15}$\\
$(6,1)_{4/3}$ &$8.7\times 10^ {15}-2.1\times 10^ {16}$&$5.9\times 10^ {15}$\\
$(8,1)_{0}$   &$5.5\times 10^ {15}-2.3\times 10^ {16}$&$3.0\times 10^ {15}$\\
$(8,1)_{1}$   &$1.0\times 10^ {16}-2.2\times 10^ {16}$&$6.5\times 10^ {15}$\\
$(8,2)_{1/2}$ &$2.5\times 10^ {16}-1.7\times 10^ {17}$&$4.8\times 10^ {15}$\\
$(10,1)_{1}$  &$1.3\times 10^ {16}-2.2\times 10^ {16}$&$1.0\times 10^ {16}$\\
\hline
\end{tabular} 
\caption{$M_{GUT}$ and $M_X^{min}$ are presented for all the {\it good}
  multiplets in Eqs.~(\ref{su5sm})  and (\ref{trnsm}). If they are heavier
  than $M_X^{min}$, the coupling constants unify within 99.73\% confidence
  level. Note, here we have assumed that the multiplet ${\bf (8,1)_{0}}$ gets
  mass from self coupling.}    
\label{mx}
\end{table}

\subsection{Unification with doublets and triplets}

The insertion of a full multiplet of the unifying group does not change any
low energy prediction of the MSSM at one loop.
However to build a phenomenologically viable model we often
need  to introduce mass-splittings between the elements of the full
multiplet. One popular example is the famous doublet triplet splitting of a 
{\bf 5} of SU(5). The doublet is identified as the MSSM Higgs and we end up
with an extra triplet (the colored Higgs).
In fact, one extra triplet slightly lighter than $M_G$ improves unification.  

The manifestation of this issue is more prominent in a $\mathrm{SU}(3)^3$
theory. The full multiplet {\bf 27} of $\mathrm{SU}(3)^3$, used in  
trinification to extract the MSSM particles as well as to break
$\mathrm{SU}(3)^3$ to the MSSM gauge group, is the sum of three
representations. 
\beq
27=(3,\bar 3,1)+(\bar 3,1,3)+(1,3,\bar 3)
\eeq    
Usually the ${\bf (1,3,\bar 3)}$ representation is given a vev to achieve the
correct breaking and this singles out ${\bf (1,3,\bar 3)}$ from the other two
components of {\bf 27}. Eq.(\ref{trnsm}) reveals that while ${\bf (1,3,\bar 3)}$
contains three doublets, three triplets come from ${\bf (\bar 3,1,3)}$. 
Hence, a mass-split between  ${\bf (1,3,\bar 3)}$ and  ${\bf (\bar 3,1,3)}$
is roughly equivalent to having three  doublet-triplet splittings. 
 But before embarking on  details of the      
model we discuss unification with both doublets and triplets in the
desert.

Let us introduce a doublet ${\bf (1,2)_{1/2}}$ of mass $M_D$ along with a triplet
${\bf (1,3)_{-1/3}}$ at a scale $M_T$. Using the fact that the scenario of  
$M_T=M_D$ is equivalent to introducing a full {\bf 5} of SU(5), the problem of two
scales $M_T$ and $M_D$ can be reduced to only one effective parameter
$M_D/M_T$. It is straightforward to see that, having a heavier
triplet compared to the doublet is the same as the
problem of a single doublet at a scale $M_G(\frac{M_D}{M_T})$. Alternatively, 
for $M_D > M_T$, the combination of a doublet and a triplet works
as a  single triplet of mass $M_G(\frac{M_T}{M_D})$.  
Thus, the deciding factor here is the parameter $\eta=\log_{10}(M_D/M_T)$.

\begin{figure}[h]
\begin{center}
\includegraphics[width=0.75\textwidth]{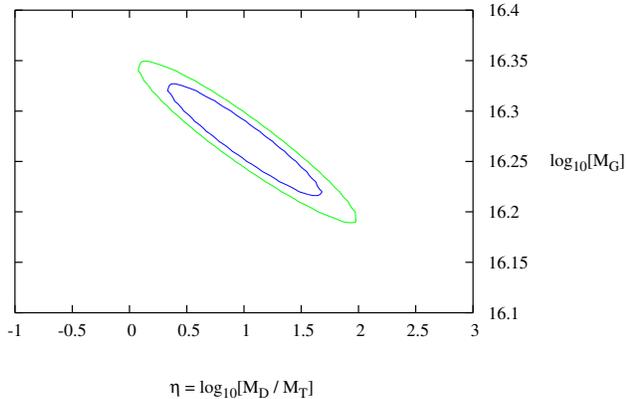}
\end{center} 
\caption{Plot showing 95\% and 99.73\% contours in the plane of $\eta$
  and $M_G$. This clearly shows that from the viewpoint of unification a
  triplet lighter than a doublet is preferred.}
\label{eta}
\end{figure}

We scan the plane of  $\eta$ and $M_G$ to predict 
the data in Eq.(\ref {input}). As shown in Fig.(\ref{eta}), present data    
clearly indicate a positive value of $\eta$ within 99.73\% confidence level.
Note that the 99.73\% ellipse in Fig.(\ref{eta}) do not touch the $\eta=0$
axis. It implies that the triplet  needs to be slightly lighter than the GUT
scale, which is in contradiction with Fig.(\ref {mssmchi}). 
In the MSSM we have two parameters $M_G$ and $\al_5$. However,
when an extra multiplet is added, we end up with three parameters
$M_X, M_G$ and $\al_5$ (in this example $M_X$ is replaced by $\eta$). 
To reduce it to two parameter plot, we determined
$\al_5$ from $M_X, M_G$ and the central value of $\al_1$. As a result, 
the ellipses shown in Fig.(\ref{eta})  shrink slightly.        

\subsection{Summary}

Before ending this section, let us summarize 
\begin{itemize}
\item We found that with MSSM particle content the gauge coupling constants
  unify at 99.73\% confidence level.
\item We have been able to identify multiplets in various GUT schemes, which
  if appearing within a certain range of mass scales, maintain and even improve
  unification. All these low dimensional {\it good} multiplets
  along with their preferred range of masses have been tabulated in
  Table~\ref{mx}.  
\item We have also been able to single out {\it bad} multiplets in different
  models. If they become light, these  representations worsen the 
  prediction of $\al_s$ compared to the MSSM.
\item We looked carefully at the decomposition of a {\bf 27} of
  $\mathrm{SU}(3)^3$ and realized that a mass-splitting among its component 
  representations is equivalent to a doublet-triplet splitting
  scenario. We found the required constraints on the mass splitting of a
  doublet and a triplet needed to preserve unification.   
\end{itemize}

\section{Constraining Trinification}

In this section we discuss the trinification scheme of Grand
Unification from the viewpoint established in the earlier section. We 
start with a short discussion of the basic idea behind trinification. Then 
we construct phenomenologically viable models in this scheme
and check whether the constraints from unification pose any threat to these
models or not. 

\subsection{Trinification in a nutshell}

Trinification is a Grand Unified theory based on the gauge group
\beq
\mathrm{G}\equiv\mathrm{SU(3)_ {C}\times SU(3)_{L}
      \times SU(3)_{R}}
\eeq 
Near the unification scale G breaks down to 
\beq
\mathrm{G} \supset \mathrm{SU(3)_ {C}
        \times SU(2)_{L} \times U(1)_{Y}}
\eeq
The group G may be extended by means of a cyclic symmetry $Z_3$. It
acts upon the three $\mathrm{SU}(3)$'s and ensures that there is only one
gauge  coupling constant.  

The lepton, quark and the anti-quark superfields transform under G as 
${\bf (1,3,\bar{3})}$, ${\bf (3,\bar{3},1)}$ and ${\bf (\bar{3},1,3)}$
 respectively. Using Eq.(\ref{trnsm}) one can designate the components as

\beq
L^a \equiv (1,3,\bar{3})^a=
   \left[ \begin{array}{ccc}
       l_c & l_1   & l_2\\
       e_c   & \nu_1 & \nu_2
   \end{array}\right]^a;\nonumber
\eeq
\beq
Q^a \equiv (3,\bar{3},1)^a=
   \left[ \begin{array}{cc}
        q & D\\
   \end{array}\right]^a;\hspace{5 mm}
Q_c^a \equiv (\bar{3},1,3)^a= 
   \left[ \begin{array}{c}
        u_c\\ d_{c_1}\\d_{c_2}
   \end{array}\right]^a  
\eeq
where $a$ is the generation index. Let us note that every generation 
contains  one  SU(2) doublet pair and an SU(3) triplet pair in addition to 
the MSSM 
particles. This is where the small demonstration with doublet and triplet in
the last section pays off. If these extra multiplets become lighter than the
GUT scale, then we recover the parameter $\eta$ defined in the last section
as $\log_{10}(M_D/M_T)$. Only here we have three $\eta$'s, one for each
generation. Combining all three of them, we redefine  
\beq
\eta \equiv \log_{10}\big[M_{D^1}M_{D^2}M_{D^3}/M_{T^1}M_{T^2}M_{T^3}\big]
\label{def:eta}
\eeq        
This parameter is going to be a useful tool in the discussion about
unification in trinification. Right now we can make a very general statement
that, if no other multiplets become light, present data indicates that a
positive value of eta is required in order to maintain unification, as was
demonstrated  previously in the last section 
(Fig.(\ref{eta})).

Usually the MSSM Higgs are obtained from a separate field $\Phi$, which has
the same quantum numbers as the lepton field $L^a$, {\it i.e.}
\beq
\Phi \equiv
   \left[ \begin{array}{ccc}
       h_u & h_{d_1} & h_{d_2}\\
       E_c   & N_1     & N_2
   \end{array}\right];
\eeq   
$\Phi$ may get a nonzero vev to break the large group G. However, it alone
cannot give the right breaking. One way to obtain the MSSM gauge
group is to introduce another field $\Phi'$ which transforms identically to
$\Phi$ under G. 
\beq
\langle\Phi\rangle =\langle\bar{\Phi}^{\dagger}\rangle
   \left[ \begin{array}{ccc}
       0 & 0 & 0\\
       0 & 0 & M
   \end{array}\right];\hspace{3 mm}
\langle\Phi'\rangle =\langle\bar{\Phi'}^{\dagger}\rangle
   \left[ \begin{array}{ccc}
       0 & 0 & 0\\
       0 & M & 0
   \end{array}\right];\hspace{3 mm}
\label{vacua}
\eeq
Each vev leaves the same $\mathrm{SU}(2)_{\mathrm{L}}$
but two different $\mathrm{SU}(2)_{\mathrm{R}}$ 
invariant. Together they give the correct breaking. We also need to
supplement the Higgs sector with its $Z_3$ partners.

The MSSM Yukawa couplings of leptons and quarks stem from the 
couplings  $g_{ab}\Phi L^a L^b$ and $\lambda_{ab} \Phi Q^a Q_c^b$. 
Extra doublets and triplets get mass from the vev of $\Phi$. 
\beq
g_{ab}\Phi L^a L^b &\rightarrow &
  g_{ab}(\langle N_2\rangle l_c^al_1^b+h_ul_2^a\nu_1^b+h_{d_1}e_c^al_2^b)
                        \nonumber \\
\lambda_{ab}\Phi Q^a Q_c^b &\rightarrow & 
 \lambda_{ab}(\langle N_2\rangle D_c^ad_{c_2}^b + 
               h_u Q^a u_c^b+h_{d_1}Q^ad_{c_1}^b) \label{decom}
\eeq
There are two more parameters which show how the MSSM Higgs $H_u$ and $H_d$
are embedded in $h_u$ and $h_{d_1}$. They are determined in the potential
involving $\Phi$ and $\Phi'$. However, Eq.(\ref{decom}) leads to two
problematic predictions irrespective of the $\Phi-\Phi'$ potential,
\beq
\frac{m_u}{m_d} = \frac{m_c}{m_s} =\frac{m_t}{m_b} \label{mass-ratio}
\eeq
\beq
\frac{M_{D^1}M_{D^2}M_{D^3}} {M_{T^1}M_{T^2}M_{T^3}}= 
   \frac{m_e m_{\mu} m_{\tau}}{m_d m_s m_b}
\label{eq:ratio}
\eeq
Eq.(\ref{mass-ratio}) contradicts the measured quark masses and 
Eq.(\ref{eq:ratio}) does not work well with unification.  
Since the quarks are heavier than the charged leptons, Eq.(\ref{eq:ratio})
suggests that the value of $\eta$ defined in Eq.(\ref{def:eta}) is negative. 
But we have seen earlier that within the present data, we need a positive
value of $\eta$ in order keep unification to the same order of magnitude as
in the MSSM. 
One way to avoid these troubles might
be to make both  $\Phi$ and $\Phi'$  couple to the matter superfields as
well as to introduce enough mixing between  $\Phi$ and
$\Phi'$ in the potential of the Higgs sector. This introduces many
parameters into the calculation and prevents us from deriving an expression
for $\eta$.

\subsection{The Higgs Sector}
 
$\mathrm{SU}(3)^3$ is broken to the MSSM gauge group in the Higgs sector.  
Recall that at least two multiplets  $\Phi$ and $\Phi'$  and their adjoints 
$\bar{\Phi}$ and $\bar{\Phi}'$ are needed to break 
$\mathrm{G} \rightarrow \mathrm{SU}(3)_{\mathrm{C}} \times
\mathrm{SU}(2)_{\mathrm{L}}\times\mathrm{U}(1)_{\mathrm{Y}}$.  
We also  introduce  additional singlet superfields $S$ and $S'$. The
superpotential involving S takes the form
\beq
W(S) =y S(\Phi\bar{\Phi}-M^2) + h S^2 + k S^3 \label{eq:vev}
\eeq
It is easy to see that W(S) does have Eq.(\ref{vacua}) as a supersymmetric
vacuum with $\langle S \rangle = 0$. $S'$ has a similar potential, but it
involves $\Phi'$ instead of $\Phi$. To make this procedure $Z_3$ invariant
we supplement $S$ with additional singlets ($S_{CL}$ and $S_{CR}$). They
couple to the $Z_3$ counterparts of $\Phi$ ($\Phi_{CL}$ and $\Phi_{CR}$)
by a potential similar to    
Eq.(\ref{eq:vev}). Note that the superpotential of the form of
Eq.(\ref{eq:vev}) has another solution where only the singlet gets a nonzero
vev to produce the minimum. 

Construction of  the rest of the Higgs potential is tricky. The MSSM Yukawa
terms come from the couplings like $\Phi L^a L^b$ and $\Phi Q^a Q_c^b$. Hence
we must find a pair of massless electroweak doublets (to be identified with
the MSSM $H_u$ and $H_d$) from $\Phi$ or $\Phi'$.  
All other extra multiplets should either be 
heavy or form  complete multiplets of the GUT group 
if the gauge coupling unification is to be preserved.

\subsubsection{First Scenario: The Minimal Model}

The term {\it minimal} employed here is in the sense of minimal GUT 
multiplet content with interactions only at the renormalizable
level. The model is built with only two Higgs multiplets which are 
needed to produce the right breaking. 

The straightforward way is to write down all the cubic couplings
involving $\Phi,\Phi', \bar{\Phi}$ and $\bar{\Phi}'$ (viz. $\Phi^3,\Phi'^3$
etc.). One can check that this
results in a light  doublet pair $H$ and $\bar{H}$ having the quantum
numbers of $H_d$ and  $H_u$ respectively. However, 
$\bar{H}$ resides completely inside  $\bar{\Phi}$ and $\bar{\Phi}'$, which 
cannot  couple to $Q^aQ_c^b$ at the
renormalizable level to produce the MSSM Yukawa couplings and  
fails to explain high top mass.

Clearly, not only do we need a massless electroweak doublet pair but they
also need to be embedded in  $\Phi$ and $\Phi'$ to predict the correct
phenomenology.  This can be achieved
if we forbid all cubic terms that involve  $\Phi$ in the Higgs sector of the
superpotential. 
This implies that both $h_u$ and $h_{d_1}$ inside  $\Phi$ 
remain massless. Note that they
have same quantum numbers as $H_u$ and $H_d$ respectively. 

This model has been proposed earlier ~\cite{Dvali:1994wj}\cite{Dvali:1994vj}
\cite{Dvali:1996fc}. 
The authors have imposed a set of discrete symmetries to
produce the model and shown that these symmetries can also be used to 
prohibit dangerous D=5 proton decay. These symmetries and their implications 
in trinification  in  detail can be found in the references.

{\bf Gauge-coupling unification in Minimal Trinification }\\
We find that this model results in a doublet pair extra to the MSSM
particle content, {\it bad} multiplets according to Eq.(\ref{trnsm}).  
From now on we will be
referring  to these extra doublets as  $H_E$ and $\bar{H}_E$. Within the 
{\it minimal} scheme they are as light as the SUSY scale. We must find
compensatory effects from extra {\bf good} multiplets. 

There are two sources of such effects: (a) the doublet-triplet splitting in
the matter sector ({\it i.e.} $\eta$ as defined in Eq.(\ref{def:eta})) and (b)
the $Z_3$ partners of $\Phi$ and $\Phi'$.

Assuming that the contributions of the light doublets $H_E$ and  $\bar{H}_E$
are compensated by the doublet-triplet splitting in the matter sector, we
calculated the size of the splitting ($\eta$). We find that a large positive 
$\eta$ ($\sim 13-14$) is necessary to produce the desired result.    
However, as we already
have mentioned,  $\eta$ is generically negative. Note that the
definition of $\eta$ (Eq.(\ref{def:eta})) involves the logarithm
to the base 10. Hence large $\eta$ implies a huge mass hierarchy between the
extra doublet and triplet in every generation. By suitably choosing
parameters  small positive $\eta$ can be generated. However, designing even
$\eta \sim 4$ requires fine tuning. 

Hence only if the $Z_3$ partners of the Higgs multiplets 
({\it i.e.} $\Phi_{CL},\Phi_{CR}$)
become light can unification be restored. Unlike  $\Phi$, both  $\Phi_{CL}$
and $\Phi_{CR}$ get masses from the singlet vev $\langle
S_{CL} \rangle = \langle S_{CR} \rangle$ (equality is because of $Z_3$). We
find that if all these multiplets
($\Phi_{CL}$, $\Phi_{CR}$, $\Phi_{CL}'$, $\Phi_{CR}'$) obtain masses $M_c 
\simeq (8\times 10^{13}-2\times 10^{14})\mathrm{GeV}$, the gauge couplings unify at 
$M_G\simeq 2.5\times 10^{17}\mathrm{GeV}$.      

{\bf Proton decay in Minimal Trinification}\\
The cyclic $Z_3$ symmetry introduces baryon number violation by operators of
dimension 6. To see this, note that the full $Z_3$ invariant Yukawa couplings
are  
\beq
g_{ab}(\Phi L^a L^b+\Phi_{CL} Q^a Q^b+\Phi_{CR} Q^a_c Q^b_c)
  + \lambda_{ab}(\Phi Q^a Q_c^b+\Phi_{CL} Q^a_c L^b+\Phi_{CR} L^a Q^b)
\eeq 
$\Phi_{CL}$ and $\Phi_{CR}$ contain triplets (Eq.(\ref{trnsm})) and the
scalar components of these fields can mediate proton decay via D=6
operators. Note that the coupling $\Phi L^a L^b$ gives leptons their masses
and $\Phi_{CL} Q^a Q^b$ 
produces a quark-quark-Higgs triplet vertex. Because of $Z_3$, these two
different operators have the same coupling constant. Similarly, the strength
of the lepton-quark-Higgs triplet vertex (from the operator 
$\Phi_{CL} Q^a_c L^b$) and the quark masses (from the operator 
$\Phi Q^a Q_c^b$) are related.  
As an example, we estimate of proton  lifetime from 
$p \rightarrow K^+\bar{\nu}$ 
\beq
\Gamma(p \rightarrow K^+\bar{\nu}) = \frac{1}{\tau_p} \sim
   \frac{g_{11}^2\lambda_{22}^2}{16\pi^2}\frac{m_p^5}{M_c^4}   
\eeq   
$M_c\sim 10^{14}\mathrm{GeV}$ and the
values of the coupling constants depend on the exact structure of Yukawa
matrices. A naive estimation would be 
$\lambda_{22}\sim \frac{\surd{2}m_c}{v_u}$ and 
$g_{11}\sim \frac{\surd{2}m_e}{v_d}$. For $\tan\be=10$, we find that 
$\tau_p \sim 10^{39}$ years. This estimation seems to be quite safe from
recent results from Super-Kamiokande ($\tau_p \geq 2.3\times 10^{33}$ years)
~\cite{Kobayashi:SUSY04}.
However, a detailed study of models of Yukawa matrices
reveals that proton lifetime can  be far lower than $10^{39}$ years. 
To support this claim we give examples of few models where the lifetime may
substantially be brought down.
\begin{itemize}
 \item We oversimplified when we estimated $\lambda_{22}$ and
 $g_{11}$. Leptons get mass from  the couplings $g_{ab}\Phi L^a L^b$ and
 $g'_{ab}\Phi' L^a L^b$. Now imagine that the MSSM Higgs $H_d$ is mostly
 contained in $\Phi'$ but the elements in the first generation of $g$ is much
 bigger than that of $g'$ so that electron mass is mostly coming from the
 coupling $g_{ab}\Phi L^a L^b$. In that case, $g_{11}$ can be as big as
 1. Thus bringing down the proton lifetime to $10^{30}$ years.    
 \item  More subtle examples come from the understanding that flavor basis
 for quarks and leptons are not related in trinification. Quark and lepton
 masses are related to the matrices $\lambda$ and $g$ respectively. To
 simplify the scenario, let us assume that both of them are diagonalized in
 the same basis. However, they do not need to have same pattern of hierarchy
 in the matrix elements. In particular, if  $\lambda$ and $g$ have inverted
 hierarchies with respect to each other, then the first generation of quarks
 (lightest) are related to the third generation of leptons (heaviest). In
 that case proton decay is accelerated and following the crude method of
 estimations, we used earlier, we find   $\tau_p \sim 10^{32}$ years.  
\end{itemize}

The proton lifetime in trinification is model
dependent. In minimal trinification the Higgs triplets are predicted to be
light. We find that there are specific models in which protons are predicted
to decay with an observable lifetime.
The additional pair of Higgs doublets may also produce large FCNC effects, 
suppressed by its mass. Turning these arguments around, we
can use the experimental bounds on proton decay as  well as on FCNC as model
building constraints while designing the Yukawa matrices in minimal trinification.

{\bf How to avoid troublesome extra light doublets ?}\\
We must extend the model beyond the minimal scheme. This can be achieved in two
ways. In the next scenario we will keep the particle content minimal, but
introduce higher dimensional operators, which  make all extra multiplets
heavy. Lastly, we will show another approach where  new multiplets
will be added to the Higgs sector.

\subsubsection{Second Scenario: The Minimal Model + New Operators}
In the last scenario, we found that we need to forbid the trilinear
couplings ($\Phi^3, \Phi^2\Phi'$ etc.) in order to generate a high top
mass. On the other hand, this results in an extra massless electroweak 
doublet pair.

To give masses to this doublet pair we introduce 
higher dimensional operators that involve $\Phi$ and $\Phi'$.
No trilinear coupling involving $\Phi$ is employed.
Following is one example of such a potential.    
\beq
W(\Phi,\Phi') &=& 
   gS_1(\Phi\bar{\Phi}-M^2)+  g'S'_1(\Phi'\bar{\Phi}'-M^2)
       + \alpha \Phi'^3+\beta \bar{\Phi}\bar{\Phi}'^3 \nonumber\\
       & & +\frac{\gamma_1}{\Lambda}(\Phi'\bar{\Phi}\Phi\bar{\Phi}')
          +\frac{\gamma_2}{\Lambda}(\Phi\bar{\Phi}\Phi'\bar{\Phi})
              +\frac{\gamma_3}{\Lambda}(\Phi'\bar{\Phi}'\Phi'\bar{\Phi})
\eeq
This potential keeps two combinations of
$h_{d_1},h_{d_2}$ and $h_{d_1}'$, as well as  one combination of
$\bar{h}_{d_2}$ and  
$\bar{h}_{d_1}'$ massless along with $h_u$. All other doublets become
heavy. We recover the MSSM Higgs as $h_u$ and one combination of 
$h_{d_1},h_{d_2}$ and $h_{d_1}'$. The other two degrees of freedom are being
eaten  by the gauge superfields. A similar mechanism follows for the fields
with quantum numbers ${\bf (1,1)_{\pm 1}}$. While $E_c'$ and $\bar{E}_c$ get mass from
the potential, $E,\bar{E}'$ remain light and are eaten fields. 

Not only are the MSSM Higgs fields obtained from  $\Phi$ and $\Phi'$, there
is enough mixing to avoid  Eq.(\ref{mass-ratio}). Relying on
higher-dimensional operators, though, results in one doublet being slightly
light. If 
the cut-off scale $\Lambda$ is assumed to be $M_{Pl}$, we end up with a
doublet of mass $\sim 10^{14}$GeV. However this effect is easily compensated if
we make $Z_3$ partners of $\Phi$ and $\Phi'$ slightly light. Note, for 
$M_C \sim 10^{16}$ years proton becomes extremely stable.

\subsubsection{Third Scenario: The Minimal Model + New Multiplets}
An alternative approach is to enrich the Higgs sector by adding new multiplets. 
Trinification models
with one extra multiplet ~\cite{Dvali:1996fc} and  with several different
multiplets ~\cite{Lazarides:1994uw}\cite{Lazarides:1995px} 
have been proposed earlier. For completeness we mention  here how the problem
of the minimal model may be resolved by adding one
extra Higgs  superfield pair $\Phi''-\bar{\Phi}''$ with its $Z_3$
conjugates. 
None of the new multiplets have vevs.
$\Phi'',\bar{\Phi}''$ are given a mass term $M\Phi''\bar{\Phi}''$ as well as
a mixing 
with  $\bar{\Phi},\bar{\Phi}'$ by $f\bar{\Phi}\bar{\Phi}'\bar{\Phi}''$. This
rotates the doublet ${\bf (1,2)_{+1/2}}$ from $\bar{\Phi}$ and $\bar{\Phi}'$ to  
$\Phi''$. We also need additional Yukawa couplings 
$g_{ab}''\Phi'' L^a L^b$ and $\lambda_{ab}'' \Phi'' Q^a Q_c^b$ to give masses to
up-type quarks. All extra particles in the Higgs sector are at the GUT scale.
Thus proton decay is delayed beyond the reach of experiments.
We also have a doublet and a triplet in the matter sector in each generation 
that may remain light, as their masses are related to the Yukawa couplings.
However, for $\eta \sim 0$ each doublet-triplet pair forms a 5 of SU(5) and
do not change any low energy prediction at one loop.

\section{Conclusion}

$\mathrm{SU}(3)^3$ gives an alternative scenario of Grand Unification. 
There are also some indications from string theory that MSSM may be embedded
in a $\mathrm{SU}(3)^3$ based group ~\cite{Kim:2003te}. Further, trinification
is probably the safest GUT as far as proton decay is concerned.

We started out this paper with the issue of the effects of intermediate
mass scales on gauge coupling unification. We predicted masses of different
particles.
We investigated trinification with these results. We constructed
the {\it minimal} model and found that it produces the right
phenomenology. However, the constraint of unification predicts light colored
Higgs which resulted in important proton decay with an experimentally
accessible lifetime.   
We also proposed a simple extension with minimal particle content but with
higher dimensional operators. It produces the right phenomenology and 
keeps unification automatic with
proton decay delayed beyond the reach of next generation experiments.

\section*{Acknowledgments}
I am  grateful to Martin Schmaltz for numerous valuable discussions and
encouragement and for comments on the manuscript. This work was supported by
the DOE Grants DE-FG02-91ER40676 and DE-FG02-90ER40560 .

\end{document}